\documentclass[pra,aps,final,twocolumn,nofootinbib,showpacs]{revtex4-1}
\usepackage{epsfig}
\usepackage{latexsym}
\usepackage{xspace}
\usepackage{hyperref}
\usepackage[latin2]{inputenc}
\usepackage{indentfirst}
\usepackage{enumerate}
\usepackage{color}
\usepackage{colordvi}
\usepackage{amsmath}
\usepackage{amssymb}
\usepackage[english]{babel}
\usepackage{url}

\def\be{\begin{equation}}
\def\ee{\end{equation}}
\def\bea{\begin{eqnarray}}
\def\eea{\end{eqnarray}}
\def\l{\label}

\def\d{\mbox{d}}

\def\siml{\;\hbox{\kern.1em \lower.7ex \hbox{$\sim$} \kern-1.12em
 \raise.5ex \hbox{$<$} \kern.1em}}
\def\simg{\;\hbox{\kern.1em \lower.7ex \hbox{$\sim$} \kern-1.12em
 \raise.5ex \hbox{$>$} \kern.1em}}

\def\siml{\;\hbox{\kern.1em \lower.7ex \hbox{$\sim$} \kern-1.12em
 \raise.5ex \hbox{$<$} \kern.1em}}
\def\simg{\;\hbox{\kern.1em \lower.7ex \hbox{$\sim$} \kern-1.12em
 \raise.5ex \hbox{$>$} \kern.1em}}

\begin{document}

\title{Quantum statistics effects near the critical point in systems with
  different inter-particle interactions
  }

\author{ S.N.\ Fedotkin, A.G.\ Magner, and U.V.\ Grygoriev}
\affiliation{Institute for Nuclear Research NASU, 03680 Kiev, Ukraine}

\begin{abstract}
      Equation of state with quantum statistics corrections
      is derived for systems of the
      Fermi and Bose particles by using their van der Waals
       (vdW)
       and effective  density-dependent
      Skyrme mean-field interactions.
      First few orders of these corrections over the
      small quantum statistics
      parameter,
    $\varepsilon \approx \hbar^3 n(mT)^{-3/2}g^{-1}$,  where
    $n$ and $T$ are the particle number density and temperature,
    $m$ and $g$ the
    mass and degeneracy factor of particles, are analytically obtained.
    For interacting system of  nucleon and $\alpha$ - particles,
    a small impurity of
    $\alpha$ - particles to a nucleon system at leading first order
    in both $\alpha $-
    particle and nucleon small parameters $\varepsilon$ 
    does not change much the basic results for the symmetric nuclear matter in
    the quantum vdW 
    consideration. Our approximate analytical results for
    the quantum vdW and 
    Skyrme mean-field approaches
        are in a
    good agreement with
    accurate numerical calculations.

\end{abstract}
%
%\pacs{24.10.Pa, 25.75.-q, 21.65.Mn}

\maketitle

\section{Introduction}
\label{introd}

A system
of interacting hadrons, first of all, nuclear matter 
is the attractive subject 
\cite{nm-1,nm-2,nm-3,nm-4,nm-5,nm-6,nm-7,nm-8,nm-9,nm-10,GK99,nm-11}.
Realistic versions of the nuclear matter equation of state includes both
the attractive and repulsive forces
between
particles.
Thermodynamical behavior of this matter
leads to the liquid-gas first-order phase transition
which ends at the critical point.
Experimentally, a presence of the liquid-gas phase transition
in nuclear matter was  reported and then analyzed
in numerous papers
(see, e.g., Refs.~\cite{ex-1,ex-2,ex-3,ex-4,ex-5,ex-5a,ex-6}).

Recently, the proposed  van der Waals (vdW) equation of state
accounting for
the quantum statistics
was used to describe the properties of
hadronic matter \cite{marik} and was extended also
  to 
multicomponent systems \cite{vova}.
Many works 
    have presented the extensions
of the phase-transition theory to the effective density-dependent Skyrme
forces in terms of the potential density \cite{AV15,satarov0,satarov}; see also
reviews \cite{La81}. They are
especially helpful for the description of the Bose-condensate in
bosonic systems \cite{satarov}. 
Starting from the pioneer works of Skyrme
(Ref.~\cite{Sk56}) and famous Skyrme self-consistent Hartree-Fock
calculations by Vautherin and Brink (Ref.~\cite{VB72}), these forces become very
popular in nuclear physics and  astrophysics; see, e.g., review
    articles
    \cite{La81,La16}.
    In different systems of hadrons, the critical points, including the Bose condensate, for 
  the classical and quantum approaches based on the vdW and Skyrme mean-field forces were studied in
 Refs.~\cite{vova,roma,satarov0,vova1,AMS19,satarov,roma1,satarov1,SBSPV,St21-1,St21-2,KSSG21}.

The role and size of the quantum statistics effects
were analytically studied for nuclear matter, also
for pure neutron and pure $\alpha$-particle matter in Ref.~\cite{FMG19}.
 In this approximation, the dependence
of 
critical point parameters on
the particle mass $m$,
degeneracy factor $g$, and the
vdW inter-particle interaction parameters $a$ and $b$
 was described well for
each of these systems.
Our consideration
was restricted to small temperatures,
$T \siml 30$~MeV, and not too large
particle densities.  Within these restrictions, the
number of nucleons
becomes a conserved number,
and the chemical potential of
such systems
was determined by the particle number.
An extension to the fully relativistic hadron resonances in a gas
formulation with
vdW interactions between
baryons and between antibaryons was considered in
Ref.~\cite{VGS-17}.
We do not include the Coulomb forces
and make no differences between protons and neutrons
(both these particles are referred as nucleons).
In addition, under these restrictions the
non-relativistic treatment becomes very accurate
and is adopted in our studies.

In the present work we are going to apply
the same analytical method as presented in Ref.~\cite{FMG19}
for systems of nucleons and $\alpha$-particles but with
another inter-particle interaction
in terms of the density-dependent effective Skyrme potential.
This method will be applied also
to a mixed two-component system of nucleons and $\alpha$- particles.
Another attractive subject of  the application of
our analytical results to analysis of the
particle number fluctuations near the critical point of
the nuclear matter (see, e.g., Ref.~\cite{roma}, and recent Ref.~\cite{FMG20})
    will be studied in a separate forthcoming work.

The paper is organized as the following.
In Sec.~\ref{sec-2} we consider the ideal quantum gases introducing the small quantum statistics
parameter related to the de Broglie wave length.
In Sec.~\ref{sec-3},
the quantum statistics corrections of the perturbation expansion over this parameter including
the inter-particle interaction  near the critical point
    are presented
on the basis of
 the vdW model.
 Section ~\ref{sec-4} is devoted to the
extension of our analytical results
to those with using the 
effective Skyrme 
 potential.
In Sec.~\ref{sec-5},
the
quantum statistics effects
near the critical point are studied for a mixed system of the
 isotopically symmetric nuclear matter with a small impurity of 
$\alpha$-particles. The results of our calculations are
discussed in Sec.~\ref{sec-disc},  and
are summarized in
Sec.~\ref{sec-sum}.
Some details of our derivations are presented in
 Appendix.

\section{Ideal quantum gases and quantum statistics parameter}\label{sec-2}

The pressure $P_i(T,\mu)$ for the
system of 
particles (e.g., $i=N$ for nucleons, $i=\alpha$ for $\alpha$ particles)
plays the role of the thermodynamical potential in
the grand canonical ensemble (GCE)
where temperature
$T$ and chemical potential $\mu$ are independent variables \cite{LLv5}.
The particle number density
$n_i(T,\mu)$, entropy density $s_i(T,\mu)$, and energy density
$\mathcal{E}_i(T,\mu)$ are given as
\be\label{term}
  n_i=\left(\frac{\partial P_i}{\partial \mu}\right)_T,~
  s_i=\left(\frac{\partial P_i}{\partial T}\right)_\mu,~
  \mathcal{E}_i= Ts_i+\mu n_i-P_i~.
  \ee
In the thermodynamic limit $V\rightarrow \infty$ considered in the present
paper all intensive thermodynamical  functions -- $P$, $n$, $s$,
and 
$\mathcal{E}$ --
depend on $T$ and $\mu$,
 rather than on the system volume $V$, see for instance
 Ref.~\cite{BG-08}.
 We start with the
GCE expressions, $\sum_iP^{\rm id}_i(T,\mu)$, for the pressure $P^{\rm id}(T,\mu)$
and particle number density, $n^{\rm id}(T,\mu)=\sum_in^{\rm id}_i(T,\mu)$,
 for the ideal
non-relativistic
quantum gas \cite{G,LLv5},
\bea\l{Pid}
& P^{\rm id}_i=\frac13 g_i\int \frac{d {\bf p}}{(2\pi \hbar)^3}\frac{p^2}{m_i}
  \left[\exp\left( \frac{p^2}{2m_iT} -\frac{\mu}{T }\right) -
    \theta_i\right]^{-1},\\
& n^{\rm id}_i=g_i\int \frac{d {\bf p}}{(2\pi \hbar)^3}
  \left[\exp \left( \frac{p^2}{2m_iT} -\frac{\mu}{T} \right) -
    \theta_i\right]^{-1}~,\l{nid}
\eea
where $m_i$ and $g_i$
are, respectively, the particle mass and degeneracy factor of the $i$
component. The value of $\theta_i=-1$ corresponds to the Fermi gas,
$\theta_i=1$ to the Bose gas, and $\theta_i=0$ is the Boltzmann (classical)
approximation when
effects of the quantum statistics
are neglected\footnote{The units
  with Boltzmann  constant
    $\kappa^{}_{\rm B}=1$ are used. We keep the Plank constant in the
  formulae  to illustrate
  the effects of quantum statistics, 
  but put $\hbar=h/2\pi=1$ in all
  numerical calculations.  For simplicity, we omitted here and below the
 subscript id for the ideal gas everywhere where
it will not lead to a misunderstanding. }.

Equations (\ref{Pid}) and (\ref{nid}) for the pressure $P_i^{\rm id}$ and density
$n_i^{\rm id}$,
proportional to the famous Fermi-Dirac and Bose-Einstein  integrals,
can be expressed in terms of
the fugacity,
\be\l{fuga}
z \equiv \exp(\mu/T)~,
\ee
as
\bea
 P^{\rm id}_i(T,z) &
 \equiv  \frac{g_iT}{\theta_i \lambda^3_i}\,{\rm Li}_{5/2}(\theta_i z)~,
 \l{Pid-1}\\ 
  n^{\rm id}_i(T,z) &
  \equiv
    \frac{g_i}{\theta_i \lambda^3_i}\,{\rm Li}_{3/2}(\theta_i z)
    ~.
    \l{nid-1}
    \eea
    Here, $\lambda_i$ is the de Broglie
    thermal wavelength  \cite{LLv5},
\be\l{lambdaT}
\lambda_i~\equiv ~\hbar\sqrt\frac{2 \pi}{m_iT}~,
\ee
and $\mbox{Li}_\nu$
is the polylogarithmic function of order $\nu$. 
The integral representation of the polylogarithmic
functions was used in these derivations; see Eqs.~(\ref{Pid})
and (\ref{nid}), and Refs.~\cite{Grad-Ryzhik,Li}.
It is convenient also to use the power
series for the polylogarithmic functions,
\be\l{Liexp}
     {\rm Li}_{\nu}(\theta_i z)\equiv \frac{\theta_i z}{\Gamma(\nu)}
     \int_0^{\infty}\frac{\d x~x^{\nu-1}}{\exp(x)-\theta_i z}
    = \sum_{k=1}^\infty
    \frac{(\theta_i z)^k}{k^{\nu}}~,
    \ee
    where $\Gamma(x)$ is the gamma function.
   Indexes $\nu=3/2$ and $5/2$ of these functions were used in
    Eqs.~(\ref{Pid-1}) and (\ref{nid-1}).
The values of $\mu >0$, i.e., $z>1$, are forbidden
in the ideal Bose  gas. The point
$\mu=0$ corresponds to an onset of the Bose-Einstein condensation in the system
of bosons. For fermions, any values of $\mu$
are possible, i.e., integrals (\ref{Pid}) and (\ref{nid})
 [see also Eq.~(\ref{Liexp})] exist for
$\theta_i=-1$ at all real values of $\mu$.
 The power series
 [see Eqs.~(\ref{Pid-1}) and (\ref{nid-1})
      with Eq.~(\ref{Liexp})]
 is
 obviously convergent at
$z < 1$ ($z>0$)
 (see, e.g., Ref.~\cite{Li}). For the Fermi statistics
at 
  $z \simg 1$,  the integral
representation of the  corresponding polylogarithmic function
[see Eq.~(\ref{Liexp})] in
Eqs.~(\ref{Pid-1}) and (\ref{nid-1})
can be used 
    (see Ref.~\cite{Grad-Ryzhik}).
Particularly,
at $z\rightarrow \infty$ one can use the asymptotic Sommerfeld expansion of the
$\mbox{Li}_{\nu}(-z)$ functions over $1/\mbox{ln}^2|z|$;
 see Ref.~\cite{brack}.

For the nucleon gas we take $m^{}_N \cong 938$~MeV neglecting
a small difference between proton and neutron masses.
The degeneracy factor is then  $g^{}_N=4$ which takes into account
two spin and two isospin
states of nucleon.
For the ideal Bose gas of $\alpha$-nuclei,
one has $g_\alpha=1$ and $m_\alpha\cong 3727$~MeV.

At $z\ll 1$,  only one
term, $k=1$,  in series,  Eq.~ (\ref{Liexp}),  is sufficient to use 
in Eqs.~(\ref{Pid-1}) and (\ref{nid-1}) which leads
to the classical ideal gas relationship
$P_i=n_i\,T$~.
Note that this 
result 
follows automatically from
Eqs.~(\ref{Pid}) and (\ref{nid})
at $\theta_i=0~$.
The classical Boltzmann approximation
at $z\ll 1$
is valid for large $T$  and/or 
small $n$ region of the $n$-$T$ plane.
In fact, at very small $n_i$, one observes $z<1$ at small $T$ too.

%%%%% FIG.1a FUGACITY.
\begin{figure}
\begin{center}
\includegraphics[width=7.0cm,clip]{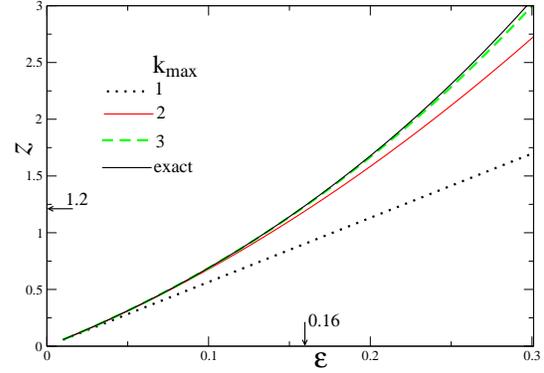}
\end{center}

\caption{
  Fugacities $z$ [Eq.~(\ref{fuga})] as functions of the
  quantum statistics parameter $\varepsilon$ [Eq.~(\ref{eps})]
  for its small values
  where one finds the critical points
   $\varepsilon_c =0.15-0.16$ [Eq.~(\ref{eps}) at $T=T_c$ and $n=n_c$;
        see Ref.~\cite{FMG19} and Table \ref{table-1} below for
  nuclear matter]. Solid black curve shows the exact fugacity
  $z(\varepsilon)$,
      found by inverting Eq.~(\ref{nid-1}) and using
  Eq.~(\ref{eps}) at $\theta=-1$,
  and
  $k_{\rm max}$ is the maximal power of cut-off series for  the polylogarithm
  $\mbox{Li}_{3/2}(-z)$; see Eq.~(\ref{Liexp}).
  Arrows show approximately the
      maximal critical values $\varepsilon_c$ and corresponding $z_c$
      under our consideration (Ref.~\cite{FMG19}).
}
\label{fig1-fe}
\end{figure}

 Inverting Eq.~(\ref{nid-1})
with respect to the fugacity, $z=z(n_i)$,  and substituting it into
Eq.~(\ref{Pid-1}), one obtains equation of state for the ideal gas through
the pressure, $P_i=P_i(T,n_i)$, for any $i$ component.
Instead of the particle number
density, $n_i$, it is convenient to introduce the
dimensionless argument, $e_i\propto n_i$, of the fugacity $z$ at a
given point of the $\mu-T$ plane:
\bea\l{eps}
&e_i \equiv
  -\theta_i\, \varepsilon^{}_{i},\qquad
  \varepsilon^{}_{i}=\frac{n_i \lambda^{3}_i}{4\sqrt2\, g_i}=D_i n_i,\nonumber\\
& D_i =
 \frac{ \hbar^3\,\pi^{3/2}}{2\,g_i\,(m_iT)^{3/2}}~.
\eea
The fugacity $z$ as function of the
  quantum statistics parameter $\varepsilon$
  for its small values
for  nuclear matter is shown in Fig.~\ref{fig1-fe}.
 Taking thus a given component $i$, e.g., for nucleon matter
($\theta_i=-1$),
 for simplicity,
 we
 omit
a subscript $i$ in
discussions of this figure.
In Fig.~\ref{fig1-fe}, the exact fugacity $z(\varepsilon)$
was obtained by multiplying  equation
(\ref{nid-1}) by the
 factor
 $\lambda^3/(4\sqrt{2}~g)$
 to get $\varepsilon=\varepsilon(z) $ and, then, inverting
 this equation with respect to $z$.

 So far, we did not use the
  series representation [Eq.~(\ref{Liexp})] for the
      polylogarithms $\mbox{Li}_\nu$
      in discussions of Fig.~\ref{fig1-fe}, in particular, for calculations of
      the solid curve ``exact''.
 Different other curves in this figure
 present the calculations for the
  maximal power, $k_{\rm max}$, in %of
  the partial sum of
  %in
  Eq.~(\ref{Liexp}) over $k$. 
    We multiply Eq.~(\ref{nid-1}) with Eq.~(\ref{Liexp}) by the same factor,
    $\lambda^3/(4\sqrt{2}~g)$,
        and use 
    cut-off of the power
  series (\ref{Liexp})
  for the polylogarithmic function
  $\mbox{Li}_{3/2}(-z)$
  at the power $k_{\rm max}$.
  As seen from this figure, one has
  the asymptotic convergence (see
  Ref.~\cite{Grad-Ryzhik})
  of $z=z(\varepsilon)$ over $k_{\rm max}$, with good convergence
   at $\varepsilon \siml 0.2$.
      Such a convergence is
  the better the
    smaller $\varepsilon$.
    Even the first-order correction is leading 
    near the critical point
    (Ref.~\cite{ex-5,ex-6,marik,FMG19}) in the region
    of $\varepsilon \ll 1$. More accurately, this region is given by
     $\varepsilon \siml \varepsilon_c \approx 0.15-0.16$,
    where $\varepsilon_c$ is the critical value of
    $\varepsilon$
      (see Eq.~(\ref{eps}) at the critical values $T=T_c$ and $n=n_c$,
       Ref.~\cite{FMG19},
      or Table \ref{table-1} below).
      This region of the variable $\varepsilon$ is related to that of
    $z \siml z_c \approx 0.8-1.2$ (see Fig.~\ref{fig1-fe}),  which
  covers well the corresponding critical value $z_c$. 
  The first (at $k_{\rm max}=2$) and, even better, second ($k_{\rm max}=3$)
  quantum statistics corrections
  improve
  the convergence.
  The cut-off sum (\ref{Liexp}) for $\mbox{Li}_{3/2}$ at the maximal power
  $k_{\rm max}=3$ practically, with the precision of lines, 
  achieves the exact result for the fugacity $z=z(\varepsilon)$
  (Fig.~\ref{fig1-fe})
  at $\varepsilon \siml 0.2$.
   In Fig.~\ref{fig1-fe}, the arrows show approximately
  the maximal values of the quantum statistics parameter
  $\varepsilon$ and corresponding fugacity $z$ for which one has still very good
  approximation by a few first-order quantum statistics corrections. 
   However, for larger
  $\varepsilon$,
 where the fugacity $z$ larger or 
about $1.5$ 
  (e.g., in the
  small temperature limit), the inversion of the cut-off sum
  [Eq.~(\ref{Liexp})] for the polylogarithmic function $\mbox{Li}_{3/2}$ fails:
  We need more
  and more terms and meet
  a divergence of the series over 
   $k$ with increasing the cut-off value of
  $k_{\rm max}$. The region of larger
  fugacity $z$ (and respectively, larger $\varepsilon$) are shown in
  Fig.~\ref{fig1-fe} 
  for the purpose of
  a contrast comparison with that of small values of $z \siml 1$,
  which are really used in our approach.
   As mentioned above,  in a
   region of a very large 
   fugacity, $z\gg 1$, 
  one has to use
  another asymptotic expansion, for instance,  over $1/\ln^2|z|$,
  as suggested by
  Zommerfeld 
  \cite{brack}.

The expansion of $z(\varepsilon)$ in powers of $\varepsilon$ is inserted
  then into Eq.~(\ref{Pid-1}).
  At small values, 
  $\varepsilon_i \ll 1$, the
  expansion of the pressure over
  powers of $\varepsilon_i$
is rapidly
 convergent.
 This expansion converges well 
 to the exact
(polylogarithmic) function results (\ref{Pid-1}) and (\ref{nid-1}).
 Its convergence is the faster the smaller $\varepsilon_i$, such that
a few first terms
 provide already a good approximation of the quantum
statistics
effects. 
 Taking  
a few first 
terms  (e.g., $k_{\rm max}=4$)
in the
    power series of Eq.~(\ref{Liexp}),
    one obtains from Eqs.~(\ref{Pid-1}) and (\ref{nid-1}) a classical gas
    result, $P_i=n_iT$,
 and the leading
first few-order corrections due to the quantum statistics effects:
\be\label{Pid-n}
P^{\rm id}_i(T,n_i)= n_i T \left[1 +
  e_i - c_2 e_i^2 - c_3 e_i^3
 + \mbox{O}(e^4_i)\right]~,
\ee
where
$c^{}_2=4[16/(9\sqrt{3})-1] \cong 0.106$~,
$c_3=4(15+9\sqrt{2}-16\sqrt{3})/3\cong 0.0201$~,
and so on.
For brevity, we will name %call
the linear and quadratic
$\varepsilon_i$-terms in
  Eq.~(\ref{Pid-n})
  as the first and second
  (order)
   quantum
  statistics corrections.

  Equation (\ref{Pid-n}) demonstrates explicitly a deviation of  the quantum
   ideal-gas pressure
   from
  its classical
  value: 
   the Fermi statistics corrections
  lead to an
  increasing of the classical pressure
  while the Bose statistics
  yields its decreasing.
  This is often interpreted \cite{LLv5} as the effective
Fermi `repulsion'
and Bose `attraction' between
particles.

\section{Quantum statistics effects with the van der Waals inter-particle
interaction
  }\label{sec-3}

Recently, the 
vdW equation of state
was extended by taking into account the effects of quantum statistics
    for nuclear matter
in Ref.~\cite{marik}.
The pressure function of the quantum vdW (QvdW) model
    for one-component
system was presented
in this paper
as
\bea\label{PQvdW}
& P(T,n)=P_{\rm id}\left[T, n_{\rm id}(T,\mu^*)\right]
- an^2~, \\
& n_{\rm id}(T,\mu^*)~=~ \frac{n}{1-bn}~,\label{nvdw}
\eea
where $P_{\rm id}$ and $n_{\rm id}$ are respectively given by  Eqs.~(\ref{Pid}) and
(\ref{nid}). 
 The modified chemical potential, $\mu^*$, is the solution of
a transcendental
equation;
see more details in Ref.~\cite{marik}
and Appendix \ref{appA}, as applied
for one-component system.
Following Ref.~\cite{FMG19}, we introduce
 a small quantum statistics
parameter $\delta$ of
expansion of the pressure $P(T,n)$ accounting for the vdW interaction
in terms of
parameters $a$ and $b$,
\be\label{del}
\delta  \equiv -\frac{\theta \varepsilon}{1-bn}=
  -\theta\, \frac{ \hbar^3\,\pi^{3/2}~n}{2\,g\,(1-bn)(mT)^{3/2}}~,
  \ee
 where $\theta$ was defined above for different statistics.
 Both first and second quantum statistics corrections over $\delta$
    to the
 vdW model
will be  presented below.

Expanding the pressure component
$P_{\rm id}\left[T, n_{\rm id}(T,\mu^*)\right]$
in  Eq.~(\ref{PQvdW})
over the small parameter $\delta$ [Eq.~(\ref{del})],
  and using Eqs.~(\ref{Pid-n}) and (\ref{nvdw}),
one obtains
\be\label{Pvdw-n}
 P(T,n) = \frac{nT}{1-bn}\left[1+\delta
  -c^{}_2 \delta^2+
    \mbox{O}\left(\delta ^3\right) \right]
   -a\,n^2,
   \ee
   where $c^{}_2$ is the same small number coefficient
   as in Eq.~(\ref{Pid-n}). We proved \cite{FMG19} that
   at small $|\delta|$
  the expansion of the pressure
  over powers of $\delta$ becomes rapidly convergent
  to the exact results. Therefore,
 a few first
    terms provide
    already a good approximation.
      A new point of
our consideration is the analytical estimates
of the
quantum statistics effects, 
    and further study of convergence of the results,
including the second order in $\delta$.
     Similarly to the ideal gases, the quantum corrections in
     Eq.~(\ref{Pvdw-n})
      increase
      with the particle number density $n$ and decrease 
      with the system temperature $T$,
particle mass $m$, and degeneracy factor $g$. A new feature of
quantum statistics
effects in the system of particles with the
    vdW
interaction is the additional
factor $(1-bn)^{-1}$ in the
correction $\delta$ [Eq.~(\ref{del})].
 Thus, the
quantum statistics effects
  become stronger
 because of the repulsive
interaction between particles.

The vdW model, both in its classical form  and in its QvdW
extension (\ref{PQvdW}) and (\ref{nvdw}),
describes the first order liquid-gas
phase transition. The critical point (CP) of this transition satisfies
the following
equations \cite{LLv5}:
\be\label{CP-0}
\left(\frac{\partial P}{\partial n}\right)_T = 0~,~~~~
\left(\frac{\partial^2 P}{\partial n^2}\right)_T=0~.
\ee
Using Eq.~(\ref{Pvdw-n}) in the first
 and second approximation
over $\delta$,
one derives
from Eq.~(\ref{CP-0})
the
system of two equations
for the CP
parameters $n_c$ and $T_c$
at the same 
 corresponding order.
 Solutions of this
 system
in the same first and second order approximation
over $\delta$ have
 the form:
\bea
 & T_c^{(1)} ~ \cong
  T_c^{(0)}\left(1 - 2 \delta^{}_0  \right)
 ~,\nonumber \\
& n_c^{(1)} \cong  n^{(0)}_c\left(1 - 2\delta^{}_0  \right)
~,  \label{nc-1}
  \eea
and
\bea
 & T_c^{(2)} ~ \cong
  T_c^{(0)}\left(1 - 2 \delta^{}_0 + \frac{4}{3}\;\delta_0^2 \right)
 ~,\nonumber \\
& n_c^{(2)} \cong  n^{(0)}_c\left(1 - 2\delta^{}_0 + 4.62\;\delta_0^2 \right)
~.  \label{nc-2}
  \eea
In Eqs.~(\ref{nc-1}) and (\ref{nc-2}),
the values $ T_c^{(0)}$ and $ n_c^{(0)}$ are the CP parameters of the classical
vdW model; see
Eq.~(\ref{CP}).
 They are
defined by Eq.~(\ref{CP-0}) and the vdW equation of state
at the zero approximation [see Eq.~(\ref{Pvdw-n}) at
$\delta=0$].
The parameter $\delta^{}_0$
in Eqs.~(\ref{nc-1}) and
(\ref{nc-2}) is given by Eq.~(\ref{del}),
 taken at the CP
    of the zero-order approximation (\ref{CP}),
i.e. at $n=n_c^{(0)}$ and $T=T_c^{(0)}$.
    For simplicity,
    we present approximately
    the number 4.62 in Eq.~(\ref{nc-2}) for a cumbersome expression.
Substituting Eqs.~(\ref{nc-1}) and (\ref{nc-2})
for the results of the
 corresponding critical temperature, $T_c^{(j)}$, and density,
$n_c^{(j)}$, where $j=1$ and $2$,
into equation of state
[Eq.~(\ref{Pvdw-n})], at a given
perturbation order, one can calculate the
CP pressure $P_c^{(j)}$
at the same order.  
Notice that the temperature $ T_c^{(1)}$ and density $ n_c^{(1)}$ are decreased for Fermi and increased for Bose particles 
 with respect to $ T_c^{(0)}$ and  $ n_c^{(0)}$.

\section{
   The
  Skyrme potential model with quantum statistics corrections}\label{sec-4}

The pressure function of the quantum
Skyrme mean-field (QSMF) model
\cite{satarov}, after some transformations, can be presented
as
\be\l{PQSkyr}
   P_{sk,i}(T,n_{i})=P^{\rm id}_ {i}\left(T, n_{i}\right)
- a^{}_{sk,i}n_{i}^2 + b^{}_{sk,i}n_{i}^{\gamma +2}~,
\ee
where $P^{\rm id}_ {i}$ is given by  Eq.~(\ref{Pid});
$a^{}_{sk,i}$, $b^{}_{sk,i}$, and $\gamma$ are
parameters of the
 QSMF parametrization.
  The
 index $i$ means, e.g.,
 nucleons N or $\alpha$ particles
 ($i=\{N,\alpha\}$).
  The QSMF parameters
 are chosen by fitting
properties of
one-component nucleon-  or $\alpha$- matter at the
    temperature $T=0$.

  Within the QSMF
 model, one can
 consider the critical points for a
first-order liquid-gas phase transition
for pure nucleon ($i=N$)  or $\alpha$ ($i=\alpha$) matter, separately.
The critical point (CP) for
     the QSMF model obeys
the
same equation (\ref{CP-0}) but
 with the
   quantum Skyrme mean-field pressure,
$P_i=P_{sk,i}(T,n_{i})$ [Eq.~(\ref{PQSkyr})] for each of components
    $i$,
\be\l{CPSkyr}
\left(\frac{\partial P_{i}}{\partial n_{i}}\right)_T = 0~,~~~~
\left(\frac{\partial^2  P_{i}}{\partial n_{i}^2}\right)_T=0~.
\ee
 For
 calculations of the first-order quantum statistics
 corrections over the small parameter $|e_i|$ %$e_i$
[see Eq.~(\ref{eps})]
to the
 QSMF pressure $P_{sk,i}(T,n_{i})$ [Eq.~(\ref{PQSkyr})],
one obtains approximately from Eq.~(\ref{Pid-n}) 
the following expression for the pressure component
$ P^{\rm id}_i(T,n_i)$ 
of Eq.~(\ref{PQSkyr}):
\be\l{PSkyr1ord}
 P^{\rm id}_i(T,n_i) = n_i\,T\,\left(1 + e_{i}\right)~.
\ee
Then, the system of two equations
    [Eq.~(\ref{CPSkyr}) for a given $i$]
    for the CP density and temperature
    values, $n_{sk,c}$ and $T_{sk,c}$,
up to the same first order over $e_i$, is reduced to
\bea\l{Skyr-cp-1}
&\!\!T_{sk}(1+2e_i)\!-\!2a^{}_{sk,i}n^{}_{sk}
\!+\!(\gamma +2)b^{}_{sk,i}n_{sk}^{\gamma +1}\!=\!0~,~\nonumber\\
&2T_{sk}e_i\!-\!2a^{}_{sk,i}n^{}_{sk}
\!+\!(\gamma +2)(\gamma +1)b^{}_{sk,i}n_{sk}^{\gamma +1}\!=0\!~.
\eea
Solving the system [Eq.~(\ref{Skyr-cp-1})] of equations
    for the CP parameters, in the first-order approximation
over $e_i$,
one obtains
\bea
& T_{sk,c}^{(1)} ~ \cong
  T_{sk,c}^{(0)}\left(1 - 2 e^{}_{i,0}\right)
~,\nonumber \\
  & n_{sk,c}^{(1)} \cong  n^{(0)}_{sk,c}
  \left(1 - \frac{2e^{}_{i,0}T_{sk,c}^{(0)}}{\gamma(\gamma+1)(\gamma+2)b^{}_{sk,i}\;
    \left[n^{(0)}_{sk,c}\right]^{\gamma+1}} \right)
~.  \label{SkyrCP-1}
  \eea
In Eq.~(\ref{SkyrCP-1}), the
temperature  $T_{sk,c}^{(0)}$ and density $n_{sk,c}^{(0)}$ are the
solutions
of equations
[see Eq.~(\ref{Skyr-cp-1})]
at zero perturbation order, $e_i=0$, 
\bea\l{SkyrCP-0}
& T_{sk,c}^{(0)}=\frac{2\gamma a^{}_{sk,i}}{\gamma+1}\left[\frac{2a^{}_{sk,i}}{b^{}_{sk,i}(\gamma+1)(\gamma+2)}\right]^{1/\gamma} ~,\nonumber ~~~\\
& n_{sk,c}^{(0)}=\left[\frac{2a^{}_{sk,i}}{b^{}_{sk,i}(\gamma+1)(\gamma+2)}\right]^{1/\gamma}~;
\eea
see also Ref.~\cite{satarov0} where another Skyrme parametrization
for the critical temperature and particle number density at zero
quantum statistics
    corrections was used.
 The parameters of Skyrme parametrization, $a^{}_{sk,i}$ and $b^{}_{sk,i}$,
    and their dimensions
    are given in the captions of Tables \ref{table-2} and \ref{table-3}.
 The value $e^{}_{i,0}$ in Eq.~(\ref{SkyrCP-1})
    is defined
by Eq.~(\ref{eps})  at $T=T_{sk,c}^{(0)}$ and $n=n_{sk,c}^{(0)}$ [Eq.~(\ref{SkyrCP-0})].
 For the CP pressure at $e_i=0$, from Eqs.~(\ref{PQSkyr}),
(\ref{PSkyr1ord}) and (\ref{SkyrCP-0}) one finds
\be\label{SkyrCP-0P}
P_{sk,c}^{(0)}=n_{sk,c}^{(0)} T_{sk,c}^{(0)}
-a^{}_{sk,i}\left[n_{sk,c}^{(0)}\right]^2
+b^{}_{sk,i}\left[n_{sk,c}^{(0)}\right]^{\gamma+2}. 
\ee
The first-order pressure, $P_{sk,c}^{(1)}$, 
can be 
 straightforwardly
calculated  from Eq.~(\ref{PQSkyr}) with using
Eq.~(\ref{PSkyr1ord}) and expressions for $T_{sk,c}^{(1)}$ and
$n_{sk,c}^{(1)}$ [Eq.~(\ref{SkyrCP-1})].

\section{
   Quantum statistics effects in the QvdW
  model for $N $ and  $\alpha$ particles system}
\label{sec-5}

For the infinite system of a mixture of different Fermi and Bose
particles, e.g.,
nucleons and $\alpha$ particles, one can present a more simple model
based on the vdW forces as a continuation of Sec.~\ref{sec-3}.
For this aim, we present the results for
the pressure function of the vdW model 
with the quantum statistics
ingredients of the QvdW model
\cite{vova},
\bea\l{PQvdWM}
&P(T,n)=P^{\rm id}_N(T,\mu^\ast_N)
+ P^{\rm id}_\alpha(T,\mu^\ast_\alpha)
\nonumber\\
&-a^{}_{NN} n^2_N -2 a^{}_{N\alpha}n^{}_Nn^{}_\alpha -
a^{}_{\alpha\alpha}n^2_\alpha~,
\eea
 where $P^{\rm id}_i$ is the pressure of an ideal
 ($i=N,\alpha$)
gas [Eq.~(\ref{PNal})].
The chemical potential, $\mu_i^\ast$, in Eq.~(\ref{PQvdWM}) is
    modified, as
shown in Appendix \ref{appA}, through the transcendent
system of equations (\ref{nNals}) and (\ref{nN})
within the QvdW model in terms of the particle number
densities $n_i$.
Following Ref.~\cite{vova}, one can fix the model
inter-particle interaction parameters $a_{ij}$ and
$b_{ij}$.
Then, it is convenient to introduce
new volume-exclusion parameters, $\tilde{b}_{ij}=2 b_{ii}b_{ij}/(b_{ii}+b_{jj})$,
where
$b_{ij}=2 \pi (R_i+R_j)$, and $R_i$ is the hard-core radius for the $i$th hard-core particle of a
multicomponent system; see Refs.~\cite{GK99,vova}.
Using the ground state
properties  of the corresponding system components,
(see, e.g., Ref.~\cite{marik}), one has %by
\bea\l{ab}
&a=a^{}_{NN}=329.8\, \mbox{MeV} \cdot \mbox{fm}^3 ~,\nonumber\\
&b=b^{}_{NN}=\tilde{b}^{}_{NN}=3.35~\mbox{fm}^3 ~.
\eea
Again, these values are very close to those
found in  Refs.~\cite{marik,vova}.
Small differences appear  because of  the
non-relativistic formulation used in the
present studies.
 For
    simplicity, for other
    attractive inter-particle interaction components we will
put
\cite{vova}
\be\l{aij}
a^{}_{N\alpha}=a^{}_{\alpha N}=a_{\alpha\alpha}=0~.
\ee
 For the repulsive-interaction components
    $\tilde{b}^{}_{ij}$ of the vdW
exclusion-volume constants
we will use those of Ref.~\cite{vova}:
 \bea\l{bij}
& \tilde{b}^{}_{\alpha\alpha}=16.76~\mbox{fm}^3, \nonumber\\
 &\tilde{b}{}_{\alpha N}=13.95~\mbox{fm}^3,~~~\tilde{b}{}_{N\alpha}=
 2.85~\mbox{fm}^3~.
 \eea
 Notice that
 the system of $N$ and $\alpha$ particles was studied
 in Ref.~\cite{satarov} within the quantum
 Skyrme mean-field  model
(Sec.~\ref{sec-4}).
 However, the authors of this article
criticized the
QvdW approach
because the Bose condensation can not be described within the
QvdW model. This phenomenon is out of scope of the present study,
and will be worked out
within our analytical approach based
on the QSMF model of the previous section
in the forthcoming work.

In the Boltzmann approximation,
i.e. at $\theta=0$ in Eqs.~(\ref{Pid}) and (\ref{nid}), the
 quantum vdW model
is reduced to the classical vdW one
\cite{vova},
\be\l{vdW}
P_i=\frac{n_iT}{1-n_j\tilde{b}_{ij}} - a_{ij}n_in_j~,
\ee
where the summations over double repeated subscripts $j$ are implied.
Note that the classical
 vdW approach (\ref{vdW}) is further reduced
to the ideal
classical gas, $P_i=n_iT$, 
at $a_{ij}=0$ and $b_{ij}=0$.
At $a_{ij}=0$ and $b_{ij}=0$, the QvdW approach
turns into that of the  quantum ideal gas
 [Eqs.~(\ref{Pid}) and (\ref{nid})].

 As mentioned above, a few
first
quantum statistics corrections
of the QvdW model
will be considered.
Expanding the pressure component
$P^{\rm id}_i\left(T, \mu^\ast_i\right)$ 
 [Eq.~(\ref{PNal})]
over
 small parameters $|e^\ast_i|$
 ($i=N,\alpha$), 
given by Eq.~(\ref{eps}) with
    replacing
    $n_i$ by $n_i^\ast$, which will be used below, at the first order in $e^\ast_{i}$
one obtains
\be\l{Pvdw-nA}
 P^{\rm id}_i(T,n^\ast_i) = n^\ast_i\,T\,\left(1+e^\ast_{i}\right)~.
\ee

This expression is similar to those of
Eq.~(\ref{Pid-n}) at the first order
over $e_i$.
     As found below,
 at small $e^\ast_i$,
the expansion of the pressure
over powers of $e^\ast_i$
becomes rapidly convergent to the exact results, and
 even the first term
 provides already a good approximation.
Our results of Eqs.~(\ref{PQvdWM}) and (\ref{Pvdw-n})
 for equations of state,
     in contrast to Eq.~(\ref{Pid-n}), as
     discussed in Refs.~\cite{LLv5,BR75},
     take into account the particle interaction effects
     (cf.\ with
     Sec.~\ref{sec-3}).
      A new point of
our consideration now is the analytical estimates
of the quantum statistics
effects
in a mixed system of
interacting fermions and bosons.
Similarly to the ideal gases, the quantum statistics
corrections in
Eqs.~(\ref{PQvdWM})
and (\ref{Pvdw-n})
 are increased for Fermi or decreased for Bose particles
     with the particle number density $n_i$.
     They are decreased (or relatively, increased) with the system
     temperature $T$,
     particle mass $m_i$, and degeneracy factor $g_i$.

     As in Ref.~\cite{vova}, we introduce
     the "mass fraction" for
     the $\alpha$ - particles impurity,
\be\l{Xal}
X_\alpha= \frac{4 n_\alpha}{n_N +4 n_\alpha} \equiv
\frac{4 n_\alpha}{n}~,
\ee
where $n$ is the baryon particle-number density,
 $n=n^{}_N+4 n_{\alpha}$.
 According to the numerical solutions in Ref.~\cite{vova},
 for the parameters of Eqs.~(\ref{ab}), (\ref{aij}), and (\ref{bij}),
 the value of
 $X_\alpha$
  [Eq.~(\ref{Xal})] has  been
 approximately obtained from a thermodynamical equilibrium of our
 mixed system,
 $X_\alpha\approx 0.013$. 
  As shown in Ref.~(\cite{roma1}), the critical point 
 in a similar two-component (neutron-proton) system, as
 function of $X_i$ ($i$ is protons), converges with
 decreasing $X_i$ smoothly to the
one-component (neutron) system.
 In line of  these results,
 one may  
 assume similarly a small change of the critical point with a
 small $\alpha$ particle impurity, $X_\alpha$, mentioned above.
 Therefore, for simplicity, we will use
 below a smallness of the $\alpha $ particle
 contribution, $X_\alpha$, in our approximate CP
 calculations by Eq.~(\ref{CP-0}), which was applied in Secs.~\ref{sec-3}
 and \ref{sec-4} for one-component systems.
Taking this
estimate for a simple exemplary case,
one can easily find $n^\ast_N$ and
$n^\ast_\alpha$ from 
Eq.~(\ref{nN}). 
 Then, using Eqs.~(\ref{Xal})
and (\ref{bij}), one can present them in the following approximate form:
\be\l{nistar}
 n^\ast_N\approx\frac{r^{}_1 n}{1- b^{}_1 n},\quad
n^\ast_\alpha\approx\frac{r^{}_2 n}{1- b^{}_2 n}~,
\ee
where 
\be\l{tbest}
 r^{}_1\approx 1-X_\alpha \approx 0.987, \quad r^{}_2\approx
    \frac{X_\alpha}{4}\approx  0.0033~.
\ee
 Here, $b^{}_1$ and $b^{}_2$
 are the
 coefficients which are related approximately
 to the repulsive
 interaction
constants $\tilde{b}_{ij}$ [$i,j=N,\alpha$; see Eq.~(\ref{bij})].
These coefficients, as functions of $\tilde{b}_{ij}$,
can be evaluated as
\be\l{tbest1}
b^{}_1\approx 3.29\,\mbox{fm}^3~,
\quad b^{}_2\approx 2.81 \,\mbox{fm}^3~.
\ee
For another modified attractive-interaction  parameter
 $a_1$, one can use
\be\l{tbest2}
a^{}_1= r^2_1\, a^{}_{NN}\approx 321.3 \,\mbox{MeV}\cdot \mbox{fm}^3~.
\ee
Using  Eqs.~(\ref{PQvdW}), (\ref{Pvdw-nA}) and (\ref{nistar}), 
    for the total system pressure
$P(T,n)$
one arrives at
\bea\l{PQvdw-na}
\!\!P(T,n)=
T \frac{r^{}_1 n \left(1\!+\!\rho^{}_1\right)}{1-b^{}_1 n}+
T \frac{r^{}_2 n\left(1-\rho^{}_2\right)}{1-b^{}_2 n}-a^{}_1n^2,
\eea
where 
\be\l{delta}
\rho^{}_{1}=\frac{D_{N}r^{}_1 n}{1-b^{}_1 n}, \quad
\rho^{}_{2}=\frac{D_{\alpha}r^{}_2 n}{1-b^{}_2 n},
\ee
$D_{N}$ and $D_{\alpha}$ are the constants given by Eq.~(\ref{eps}),
$r^{}_1$ and $r^{}_2$ are given by
Eq.~(\ref{tbest}).
Note that the  expression for the  pressure,
Eq.~(\ref{PQvdw-na}), in
 the case of $r^{}_2=0$ and $r^{}_1=1$ is exactly the same as for
a pure
nuclear matter
 presented in Ref.~\cite{FMG19}
    (see Sec.~\ref{sec-3}).
 A new
     feature of the quantum statistics
effects in the system of particles with the vdW interactions is the additional
factors 
$(1-b^{}_{1}n)^{-1}$ and $(1-b^{}_{2}n)^{-1}$  in the perturbation parameters.
 Thus, the quantum statistics
 effects become stronger
  because of the repulsive interactions between particles.

\begin{figure*}
\begin{center}
 \includegraphics[width=8.0cm,clip]{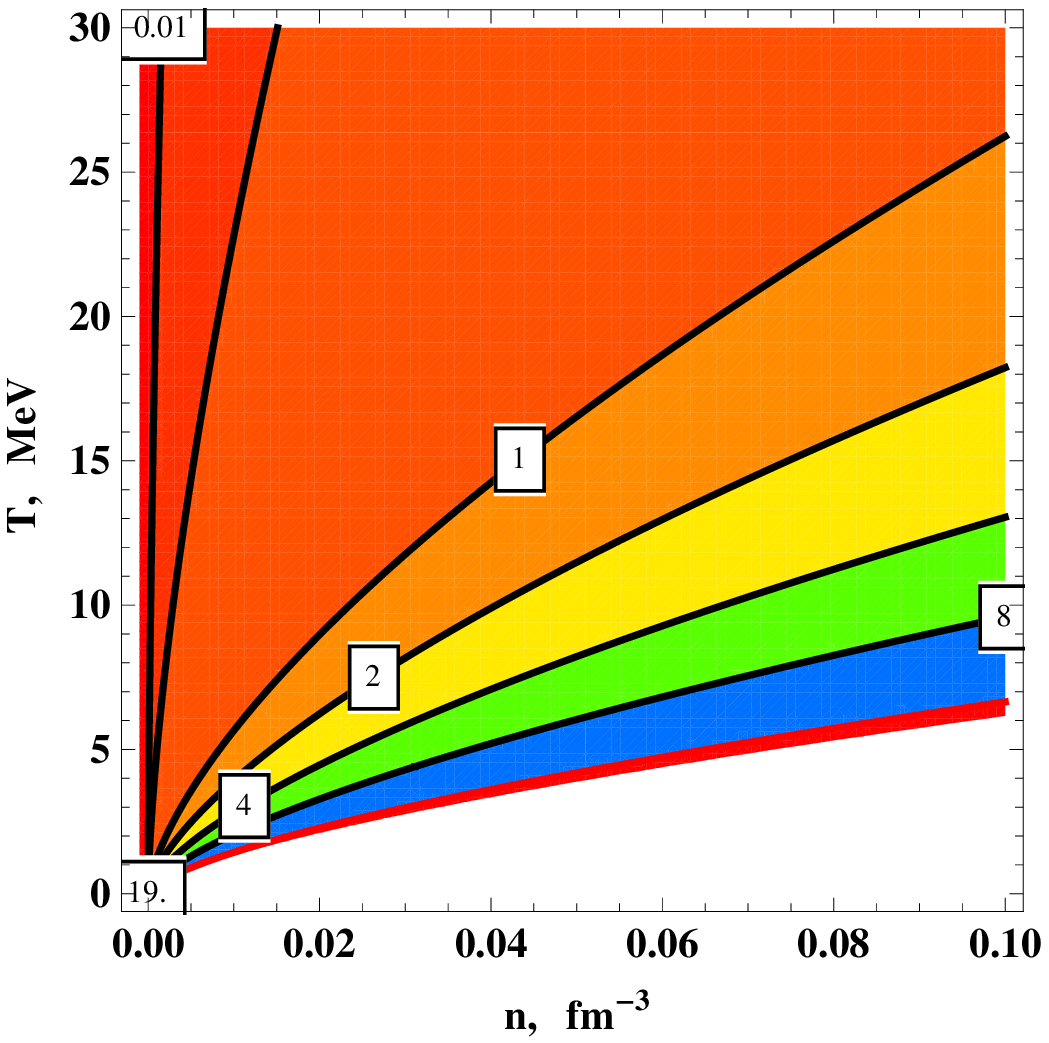}
  \includegraphics[width=8.1cm,clip]{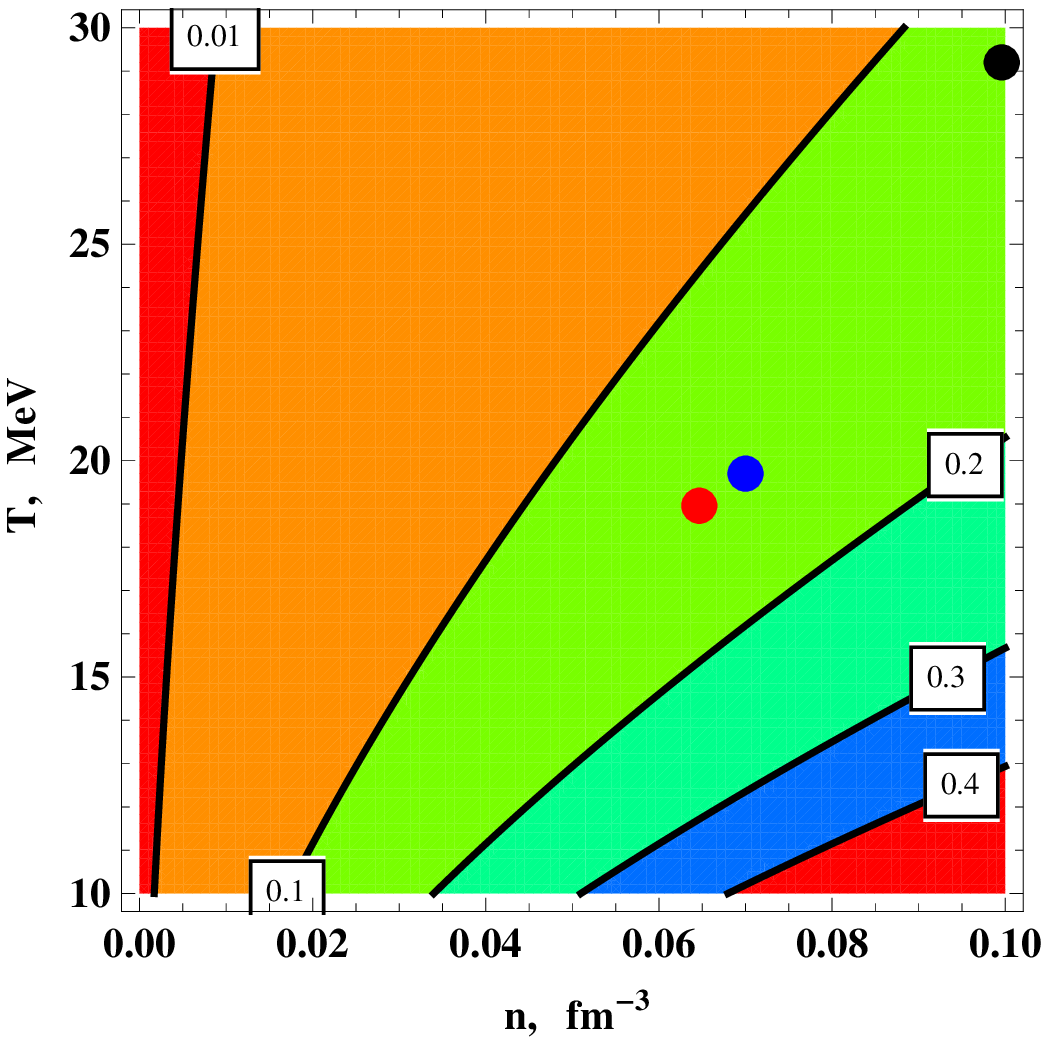}
\end{center}

\caption{
  Contour plots for the first-order fugacity $z(n,T)$ ($k_{\rm max}=2$)
  and corresponding parameter $\varepsilon(n,T)$
  for  nucleon matter in the plane of density
  $n$ and temperature $T$
  are shown in the
  left and right panels, respectively.
  The red line
  (left) shows the zero entropy line, such that the white area is
  related to a nonphysical region where the entropy of the ideal gas
  is
  negative.
The critical point for our first-order and the zero-order (standard vdW)
approximations
for nuclear matter at the parameters $a$ and $b$ [Eq.~(\ref{ab})]
are shown in right panel by
the red [see Eq.~(\ref{nc-1}), Table \ref{table-1}, and Ref.~\cite{FMG19}]
and  the black (vdW)
point, relatively. The blue point in the same plot
presents the numerical result for the critical point
 (Ref.~\cite{marik} and Table \ref{table-1}).
}
\label{fig2}
\end{figure*}
%
%%%%% FIG.3 P(v,T), P(n,T).
\begin{figure*}
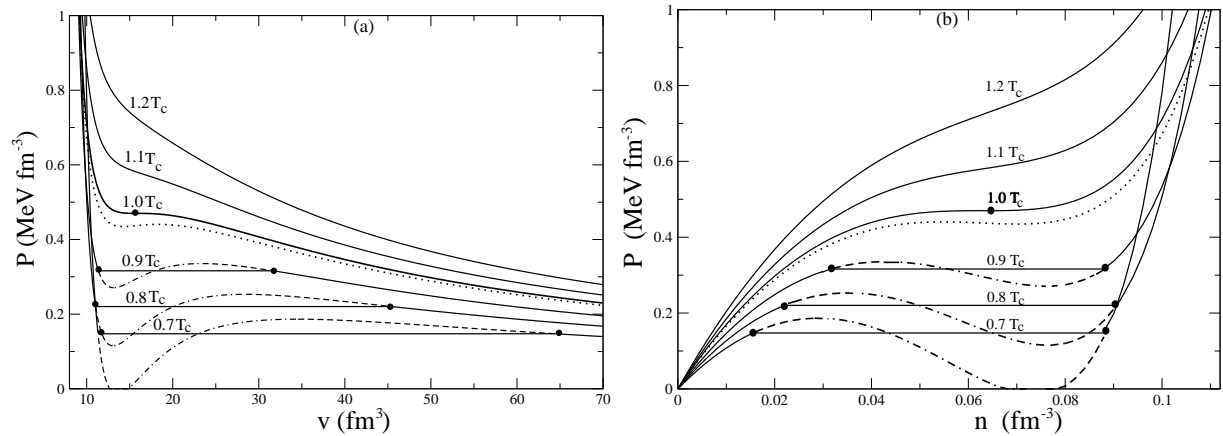

\begin{center}
\includegraphics[width=8.0cm,clip]{Fig3a_PvT-fin.eps}
\includegraphics[width=8.0cm,clip]{Fig3b-PnT-fin.eps}
\end{center}

\caption{
  Pressures $P$ as functions of the reduced volume $v$ (a)
  and
particle number density
$n$
 (b) at different temperatures $T$ (in units of the
critical value $T_c$)
at first-order in the quantum statistics expansion over $\delta$
[Eq.~(\ref{del})]
  for the simplest case of the
symmetric nucleon matter. The critical point is shown by the
close circle, see text for details.
The dotted line shows the second-order 
approximation over $\delta$;
see Ref.~\cite{FMG19}, and Eq.~(\ref{Pvdw-n}).
The horizontal lines are plotted by using the Maxwell area
law in the  panel (a) 
and correspondingly in the panel (b).
The unstable and meta-stable parts of the isothermal lines are presented
by dashed and dash-dotted lines, respectively.
    Other closed dots show schematically a binodal
    boundary for the two-phase coexistence curve.
}
\label{fig3}
\end{figure*}

The vdW model, both in its classical form (\ref{vdW}) and in its QvdW
extension
[Eqs.~(\ref{PQvdW}) and (\ref{PQvdw-na})]
describes the first-order
  liquid-gas
phase transition. As the value of $X_\alpha$, used in our derivations, is
very small, the
    critical points in the considered approach can be
    determined approximately, as mentioned above,
    by the
     same
    equations given by Eq.~(\ref {CP-0}).
    Using Eq.~(\ref{PQvdw-na}) in the first approximation
     over quantum statistics corrections,
one derives
from Eq.~(\ref{CP-0})
the
system of two equations
for the CP
parameters $n_c$ and $T_c$
at the same first order:

%%%%%%%%%%%%% TABLE 1 %%%%%%%%%%%%%%%%%%%%%%%%%%%%
\vspace{0.3cm}
\begin{table*}[pt]
\begin{center}
\begin{tabular}{|c|c|c|c|c|c|c|}
\hline
Critical point
& vdWM
& $N$ 1st-order & $N$ 2st-order & $N$ numerical  & $N+\alpha$ 1st-order  & $N+\alpha$ numerical\\
 parameters & Eq.~(\ref{CP})
& Eq.~(\ref{nc-1}) & Eq.~(\ref{nc-2}) & full QvdW & Eq.~(\ref{cp-1ab})
& full QvdW \\
\hline
$T_c$~[MeV] & ~29.2~& ~19.0~& ~20.0
& ~19.7~&~19.4~&~19.9\\
\hline
$n_c$~[fm$^{-3}$] &0.100 & 0.065 & 0.079
& 0.072 &~0.072~&~0.073\\
\hline
$P_c$~[MeV$\cdot$ fm$^{-3}$] & 1.09  & 0.48 & 0.56
& 0.52 &~0.51~&~0.56\\
\hline
\end{tabular}
\vspace{0.2cm}
\caption{{\small
    Results for the CP parameters of the van der Waals model 
        (2nd column), the symmetric nuclear
matter ($N$) ($g^{}_N=4, ~m^{}_N=938~\mbox{MeV}$, 3rd, 4th and 5th columns), and
the mixed  $N+\alpha$ matter
 ($g^{}_\alpha=1, ~m^{}_\alpha=3737~\mbox{MeV}$,
6th and 7th columns). Numerical results
obtained within the full 
 QvdW model in Refs.~\cite{marik} and \cite{satarov}
   are shown in 5th and 7th columns,
respectively.
}}
\label{table-1}
\end{center}
\end{table*}

\vspace{-0.7cm}
\bea\l{cp-1ab}
&2na^{}_1 = \frac{T r_1\,\left(1+2\rho_1\right)}{\left(1-b^{}_1n\right)^2}
+\frac{T r_2\,\left(1-2\rho_2\right)}{\left(1-b^{}_2n\right)^2}~,
\nonumber\\
    &a^{}_1 = \frac{T r_1 b_1}{\left(1-b^{}_1n\right)^3}
\left[1+\rho^{}_1\frac{(1+2b_1n)}{b^{}_1n}\right]\,\nonumber\\
  &+\frac{T r_2 b_2}{\left(1-b^{}_2n\right)^3}
\left[1-\rho^{}_2\frac{(1+2b_2n)}{b^{}_2n}\right]~.
\eea
Note that 
 Eq.~(\ref{cp-1ab}) for the CP
in
the case of $r^{}_1=1$, $r^{}_2=0$ and $b^{}_1=b^{}_{NN}$ is exactly the same as
that for a pure
nucleon
matter,  which was derived in Ref.~\cite{FMG19}
(see Sec.~\ref{sec-3}).

\section{Discussion of the results}
\l{sec-disc}

  A summary of the results for CP parameters
     are presented for the QvdW model and Skyrme mean-field
     parametrization in Figs.~\ref{fig2} and \ref{fig3}  and
     Tables \ref{table-1}-\ref{table-3}. 
     Fig.~\ref{fig2} shows
     the contour graphics in the $n-T$ plane where black lines  mean
     $z(n,T)=const$ [see Fig.~\ref{fig1-fe} and
       Eqs.~(\ref{nid-1}) and (\ref{fuga})]
      in the left and $\varepsilon(n,T)=const$ [Eq.~(\ref{eps})]
  in the right panel with the  constant values written in
  white squares. As seen from these plots, all values of
  $z \siml 1$ ($z \siml 1.2$)
  correspond to
  $\varepsilon \ll 1$ ($\varepsilon \siml 0.2$) above blue regions. 
  Therefore, together with Fig.~\ref{fig1-fe},
  this explains the reason for using the expansion
   over small parameter $\varepsilon$,
  even when
  the fugacity is of the order of one, having a little larger values.
  In particular,
  the critical points,
  obtained in Ref.~\cite{FMG19} and shown in
      Fig.~\ref{fig2} and Table \ref{table-1}, belong
  such a region.
%
%%%%%%%%%%%%% TABLE 2 %%%%%%%%%%%%%%%%%%%%%%%%%%%%
\vspace{0.3cm}
\begin{table}[pt]
\begin{center}
\begin{tabular}{|c|c|c|c|}
\hline
Critical point  & 0th-order   & 1st-order & numerical\\
parameters & Eq.~(\ref{SkyrCP-0}) & Eq.~(\ref{SkyrCP-1})  & full QSMF \\
\hline
$T_{sk,c}$~[MeV] & ~20.06~& ~15.1~
& ~15.3~~\\
\hline
$n_{sk,c}$~[fm$^{-3}$] &0.06 & 0.047
& 0.048
\\
\hline
$P_{sk,c}$~[MeV$\cdot$ fm$^{-3}$] & 0.325  & 0.194
& -
\\
\hline
\end{tabular}
\vspace{0.2cm}
\caption{{\small
Results for the CP parameters of the symmetric
nuclear matter in the quantum Skyrme mean-field (QSMF) model
($g=4,~m=938~\mbox{MeV},~\gamma =1/6,~a^{}_{sk,N}
=1.167~\mbox{GeV}\cdot\mbox{fm}^3,~b^{}_{sk,N}
=1.475~\mbox{GeV}\cdot \mbox{fm}^{3  +3\gamma}$).
    Numerical results obtained within the full QSMF model 
    in Ref.~\cite{satarov}
    are shown in 4th column.
}}
\label{table-2}
\end{center}
\end{table}

%
%%%%%%%%%%%%% TABLE 3 %%%%%%%%%%%%%%%%%%%%%%%%%%%%
\vspace{0.3cm}
\begin{table}[pt]
\begin{center}
\begin{tabular}{|c|c|c|c|}
\hline
Critical point  &   0th-order
  & 1st-order & numerical \\
parameters & Eq.~(\ref{SkyrCP-0})      & Eq.~(\ref{SkyrCP-1})  & full QSMF  \\
\hline
$T_{sk,c}$~[MeV] & ~9.667~& ~10.198~
& ~10.200~~\\
\hline
$4n_{sk,c}$~[fm$^{-3}$] &0.0353 & 0.037
& 0.037
\\
\hline
$P_{sk,c}$~[MeV$\cdot$ fm$^{-3}$] & 0.023  & 0.025
& -
\\
\hline
\end{tabular}
\vspace{0.2cm}
\caption{{\small
Results for the CP parameters of pure $\alpha$ matter
in the 
QSMF model ($g=1,~m=3727~\mbox{MeV},~\gamma =1/6,~a^{}_{sk,\alpha}
=3.831~\mbox{GeV}\cdot \mbox{fm}^3,~ b^{}_{sk,\alpha}
=6.667~\mbox{GeV}\cdot \mbox{fm}^{3  +3\gamma}$).
 Numerical results obtained within the full QSMF model
 in Ref.~\cite{satarov}
are shown in 4th column.
}}
\label{table-3}
\end{center}
\end{table}

\vspace{-0.6cm}
 Fig.~\ref{fig3} shows the isotherms of the pressure
 $P$ as function of the reduced volume $v=1/n$ 
and the
particle number density $n$ for an isotopically
symmetric nuclear matter.
The
 first- (and second-) order quantum statistics
 corrections are presented in this figure and Table \ref{table-1}.
 The critical point is shown by the
 close circle found from the 
accurate solution of
equations (\ref{CP-0})
 [see  Eq.~(\ref{nc-1})]
    for nucleon matter; see also the close red circle in Fig.~\ref{fig2}.
   The dotted line shows the second-order 
   approximation
   [see Eq.~(\ref{nc-2}) 
   and Eq.~(\ref{Pvdw-n})]
     at the same
      nuclear matter parameters.
Dotted line presents schematically a binodal
boundary for the two-phase coexistence curve in
the transition from the two- to one-phase range \cite{FMG19}.

Results for the CP parameters obtained by
Eqs.~(\ref{nc-1}) and (\ref{nc-2}),
and by solving the system
of equations, Eq.~(\ref{cp-1ab}),
are presented in Table \ref{table-1}.
These analytical results are close to the numerical
results obtained in Refs.~\cite{marik,vova}.
 For the same nucleon matter case,
 a difference of the results for the  
 vdW
[Eq.~(\ref{CP})] and
QvdW [Eqs.~(\ref{nc-1}) and (\ref{nc-2})] models
 in
Table \ref{table-1}
demonstrates
a significant role of the effects of
 quantum statistics
for the CP of the symmetric
nuclear-particle matter.
Table \ref{table-1} shows also
  a good convergence
  for the case of the mixed $N-\alpha$
  system with a small $X_\alpha$, given by Eq.~(\ref{Xal}). Even the first-order
  corrections are in good agreement with the exact numerical QvdW results;
  see Refs.~\cite{marik,vova,FMG19} and
  many other examples were recently considered in Ref.~\cite{VOV-17}.
As seen from
  Table \ref{table-1}, the quantum statistics
  effects of the $\alpha$- particle impurity
   are, in fact, small because, first of all, of too
  small relative concentration
  $X_\alpha$ of this impurity, according to Eq.~(\ref{Xal}),
  which was suggested in
  Ref.~\cite{vova}. The size of
these effects appears to
be rather different
for the case of impurity contributions
$X_\alpha \cong 1$ of the
$\alpha$ particles into the nucleon matter.

As stated above, our analysis can be applied
beyond the vdW model. In fact, similar estimates
of the quantum statistic effects
have been
done also
for one of the mean-field
models   (Ref.~\cite{AV15} and references therein) in
Sec.~\ref{sec-4}; see Tables \ref{table-2} and \ref{table-3}.
The QSMF 
calculations was performed 
  for $\gamma=1/6$ and other corresponding parameters from
  Ref.~\cite{satarov} and presented in these Tables.
  There is very good agreement between analytical
  results of calculations (\ref{SkyrCP-1})
  up to the first-order corrections over $e_i$ and
  numerical results obtained in Ref.~\cite{satarov};
  see Table \ref{table-2} for nucleon matter and Table \ref{table-3}
  for $\alpha$ matter.
  A similar good agreement was obtained with the results of Ref.~\cite{satarov},
  also for another parameter,
  $\gamma=1$, 
   which was found in  the derivations of SMF approach
  \cite{La81} from
  the original Skyrme forces \cite{Sk56,VB72}.
 Thus, 
  the first
  order over small parameter $e_i$ in expansion of the pressure
  within the quantum Skyrme mean-field approach, as well
  in the QvdW model,
  turns out to be sufficient for very good agreement with
  numerical calculations \cite{satarov} of the CP parameters.

  We should emphasize also that it is remarkable that the  results obtained
  up to the
  first-order corrections
reproduce the quantum statistics
effects
with a high accuracy (see Tables \ref{table-1}-\ref{table-3}).
The contribution of high-order (e.g., second-order)
    corrections in expansion over
    $\delta_i$ for
 the QvdW model, or over
 $e_i$ for the Skyrme-mean field parametrization,
is much
  smaller than the first-order correction,
  that
  shows
  a fast convergence
  in $\delta_i$, or $e_i$,
  by the
  first-order terms.
 Notice  that a
      smallness of the
      parameters $\delta_i$ is associated with  those of
      $\delta_i\propto e_i$.
  Therefore,  high-order
  corrections due to the quantum statistics
  effects can be neglected
  for main evaluations  of the critical point
  values.

  \section{Conclusions}
\label{sec-sum}

 We derived the critical point temperature,
particle number density, and pressure for the nucleon and $\alpha$ particle
matter within the quantum van der Waals (QvdW) and quantum Skyrme mean-field
 (QSMF) models
taking into account the Fermi and Bose statistics corrections. We found their
analytical dependence on the system parameters of particles, such as their
mass, degeneracy, and inter-particle interaction constants.
 It is shown that, in order
 to determine the equation of state and the critical
   point with
 accounting for the quantum statistics and interaction effects,
 it is sufficient to keep
 only the first term in the pressure expansion
  basically over the small quantum statistics parameter $e_i$,
  where $i$ denotes the
     baryon system under consideration.
 The specific properties of particles
 (their mass and degeneracy
factor) appear in the CP values through this small
parameter $e_i$.

  Our derivations were carried out
 for systems of Fermi or Bose particles in two cases:
 for the 
 QvdW model and the 
QSMF parametrization.
 In both cases, taking into account already the first-order terms in
 expansion of the pressure
  over $e_i$
 greatly simplifies the form of the equation of state
  and solution of this equation
 for the critical values of temperature, density, and pressure. The
 values of these critical quantities at leading first order
 turn out to be very close to those obtained in accurate
 numerical calculations 
 within the full 
 QvdW and QSMF models.
 For relatively small temperatures $T$ and/or
 large particle-number densities
 $n$,
 the quantum statistics parameter, 
 $|e_i| \propto n_i T^{-3/2}$,
becomes large. In this region of the phase
diagram, the perturbation expansion diverges and, therefore,
the 
QvdW and QSMF approaches
should be treated within the full quantum statistical
  formulation.
 However, as well known \cite{LLv5}, for the limit of small
temperatures $T$ and /or large
particle densities $n$, the vdW approach fails.
In particular,  as shown early (Ref.~\cite{satarov}),
the Bose condensation phenomenon
    should be treated within the QSMF model,
    in contrast to the vdW approach.

A simple and explicit dependence on the system parameters,
such as the particle mass $m_i$ and degeneracy factor $g_i$,
is demonstrated at the leading few
first orders of %this
expansion over $e_i$.
Such a dependence is absent within the classical van der Waals
and
Skyrme mean-field models.  The quantum statistics parameter  $e_i$,
is proportional to $m_i^{-3/2} g_i^{-1}$. Therefore,
the effects of quantum statistics  become smaller for
 more heavy
particles and/or for larger
values of their degeneracy factor.

The quantum statistics corrections to the CP parameters
of the symmetric nuclear
 matter appear to be quite significant.
  For a pure nuclear matter, the value of
     $T_c^{(0)} = 29.2$
     ~MeV in the classical vdW model
  is decreased dramatically
to the QvdW value $T_c^{(1)}=19.0$~MeV
 at the first-order approximation in the
    quantum statistics expansion.
On the other hand,
this approximate analytical  result within the first-order
 quantum-statistics approach
is already  close to the accurate
numerical value of $T_c=19.7$~MeV,
 which was obtained by
    numerical calculations within the full QvdW model.
    For the Skyrme
mean-field parametrization, the quantum statistics effect is
smaller than that for the quantum van der Waals model.
 This improves  the
foundation of the
perturbation approach used with respect to
the small parameter $e_i$.
The agreement of the first-order 
 QSMF approach
with full numerical calculations \cite{satarov}
is even
better than that within the QvdW model.
The nuclear matter value of
$T_c^{(0)}=20.6$~MeV
in the classical SMF case is decreased 
to the quantum SMF
value $T_c^{(1)}=15.1$~MeV.
 This result % which
is obviously very close to
 that of numerical calculations, $T_c=15.3$~MeV,
 obtained in Ref.~\cite{satarov}.

 The QwdW
 equation of state has been derived analytically and used to study
the quantum statistics effects
in a vicinity of the critical point of the two-component system of
nucleon and $\alpha$-particle matter.
The expressions for the pressure of the equation of state
were obtained by using the quantum statistics expansion
over the two small parameters $e^{\ast}_i$
($i=\{N,\alpha\}$) near the vdW approach.
 The trend of the critical-value corrections,
  owing to an impurity
of the
$\alpha$ particles into the nucleon system,
occurs in 
 the following direction:
the
CP parameters are  somewhat increased
as compared to those for pure nucleon system.
These analytical results are
in good agreement with those of more accurate
numerical calculations.
 A very small impurity
of the $\alpha$
particles to the nucleon matter leads to a very
small
corrections to the equation of state and to the
critical point
of the nuclear matter.

 Finally, one can conclude that our derivations within 
 the quantum van der Waals   
 and Skyrme mean-field parametrization
are straightforwardly
extended to other types of inter-particle interactions. In particular,
it is expected to be the case for a more general mean-field approach.

\begin{acknowledgments}
  We thank M.I.~Gorenstein and A.I.~ Sanzhur for fruitful discussions and
  suggestions, as well  D.V.~Anchishkin, A.~Motornenko, R.V.~Poberezhnyuk,
  and V.~Vovchenko for many useful discussions.
The work of S.N.F. and A.G.M. on the project
``Nuclear collective dynamics for high temperatures and
neutron-proton asymmetries'' was
supported in part by the Program ``Fundamental researches in high energy physics
and nuclear physics (international collaboration)''
at the Department of Nuclear Physics and Energy of the National
Academy of Sciences of Ukraine.  S.N.F., A.G.M. and U.V.G. thank
the support in part by the budget program ``Support for the
development of priority   areas of scientific researches'',
the project of the Academy
of Sciences of Ukraine (Code 6541230, No 0120U100434).
\end{acknowledgments}

\appendix

\renewcommand{\theequation}{A.\arabic{equation}}
\renewcommand{\thesubsection}{A\arabic{subsection}}
  \setcounter{equation}{0}

\section{To formulation of the QvdW model}
\l{appA}

The  constants of the  QvdW model,
$a>0$ and $b>0$, are responsible for respectively attractive and repulsive
interactions between particles.
We fix the model parameters $a$ and $b$ using the ground state
properties  of  the symmetric nuclear matter
(see, e.g., Ref.~\cite{nm}): at $T=0$ and
  $n=n_{0}= 0.16~\mbox{fm}^{-3}$, one requires
$P=0$ and the binding energy per nucleon $\varepsilon(T=0,n=n_0)/n_0=- 16$~MeV.
 From the above requirements\footnote{The multicomponent
QvdW model with different $a$ and $b$ parameters for protons and neutrons was
discussed in Refs.~\cite{vova,roma}.},  one
finds
\be\l{ab1}
a=329.8\, \mbox{MeV} \cdot \mbox{fm}^3 , \;\;\; b=3.35\, \mbox{fm}^3~.
\ee
 Therefore, in Eqs.~(\ref{nc-1}) and (\ref{nc-2}),
    one has
the CP parameters of the classical vdW model:
\bea
  & T_c^{(0)}=\frac{8a}{27b}\cong 29.2~{\rm MeV}~, ~~~
  n_c^{(0)}=\frac{1}{3b}\cong 0.100~{\rm fm}^{-3}~,\nonumber \\
& P_c^{(0)}=\frac{a}{27b^2}\cong 1.09~{\rm MeV}\cdot {\rm fm}^{-3}~. \label{CP}
\eea
They
were found from Eq.~(\ref{CP-0})
for the equation of states  in 
the case $\delta=0$.

Following the formulation of the QvdW model \cite{vova}, one can present
the equation of state
(\ref{PQvdWM}) in terms of the pressures of the ideal 
two-component gas in the grand canonical ensemble,
but in terms of the modified chemical potentials $\mu^\ast_i$
($i=\;N,\alpha$). These potentials, $\mu^\ast_i$, were defined through
the particle number densities $n_i$ by
a system
of transcendent
equations.  First, they are found as functions
    of the densities $n^\ast_i$
by solving the following equations:
\bea\l{nNals}
&n^\ast_N = n^{\rm id}_N(T,\mu^\ast_N) \equiv
\frac{2g^{}_N}{\sqrt{\pi}~\lambda^3_N}
\int_0^\infty d\eta
\frac{\eta^{1/2}}{\exp\left(\eta - \frac{\mu^\ast_N}{T}\right) +1}~,\nonumber\\
&n^\ast_\alpha = n^{\rm id}_\alpha(T,\mu^\ast_\alpha)\equiv
\frac{2g_\alpha~}{\sqrt{\pi}~\lambda^3_\alpha}
\int_0^\infty d\eta
\frac{\eta^{1/2}}{\exp\left(\eta - \frac{\mu^\ast_\alpha}{T}\right) - 1}~,
\eea
\\
\noindent
where
$n^{\rm id}_i$ is defined by Eq.~(\ref{nid}),
see more details in Refs.~\cite{marik,vova}.
Then, $n^\ast_i$ can be obtained in terms of the true
particle-number
densities $n_i$ of interacting particles system
where interaction is described by the 
 vdW exclusion-volume constants $\tilde{b}_{ij}$, Eq.~(\ref{bij}),
 from the equations: 
\bea\l{nN}
&n^{}_N=\frac{n^\ast_N\left[
    1+\left(\tilde{b}_{\alpha\alpha}-\tilde{b}^{}_{\alpha N}\right)n^\ast_\alpha\right]}{
  1+\tilde{b}^{}_{NN}n^\ast_N+\tilde{b}_{\alpha\alpha}n^\ast_\alpha +\left(\tilde{b}^{}_{NN}\tilde{b}_{\alpha\alpha}
-\tilde{b}^{}_{N\alpha}\tilde{b}^{}_{\alpha N}\right)n^\ast_Nn^\ast_\alpha}~,\nonumber\\
  &n^{}_\alpha=\frac{n^\ast_\alpha\left[
    1+\left(\tilde{b}^{}_{NN}-\tilde{b}^{}_{N \alpha}\right)n^\ast_N\right]}{
  1+\tilde{b}^{}_{NN}n^\ast_N+\tilde{b}_{\alpha\alpha}n^\ast_\alpha +\left(\tilde{b}^{}_{NN}\tilde{b}_{\alpha\alpha}
  -\tilde{b}^{}_{N\alpha}\tilde{b}^{}_{\alpha N}\right)n^\ast_Nn^\ast_\alpha}~.
  \l{nNal}
\eea

Finally, one can obtain the pressure $P^{\rm id}_i$ given by
Eq.~(\ref{Pid}) but with the modified chemical potential $\mu^\ast_i$, found from
Eqs.~(\ref{nNals}) and (\ref{nN}), as
\bea\l{PNal}
&
P^{\rm id}_N(T,\mu^\ast_N)=
\frac{4g^{}_N~T}{3\sqrt{\pi}~\lambda^3_N}
\int_0^\infty d\eta
\frac{\eta^{3/2}}{\exp\left(\eta - \frac{\mu^\ast_N}{T}\right) +1}~,\nonumber\\
&
P^{\rm id}_\alpha(T,\mu^\ast_\alpha)=
\frac{4g_\alpha~T}{3\sqrt{\pi}~\lambda^3_\alpha}
\int_0^\infty d\eta
\frac{\eta^{3/2}}{\exp\left(\eta - \frac{\mu^\ast_\alpha}{T}\right) - 1}~.
\eea
    Eqs.~(\ref{nN}) and (\ref{PNal}) are used for derivations of the equation of state
 (\ref{PQvdWM}) in the main text.

\end{document}